\documentclass[a4paper,11pt]{article}
\usepackage{pos}

\title{Path integral treatment of coherence effects 
in charmonium production in nuclear ultra-peripheral collisions}

\author*[a]{J. \'{O}bertov\'{a}}
\author[a,b]{J. Nemchik}

\affiliation[a]{Faculty of Nuclear Sciences and Physical Engineering, Czech Technical University in Prague, \\ B\v{r}ehov\'{a} 7, 115 19, Prague, Czech Republic}

\affiliation[b]{Institute of Experimental Physics SAS, Watsonova 47, 04001 Ko\v sice, Slovakia}

\emailAdd{jaroslava.obertova@fjfi.cvut.cz}
\emailAdd{jan.nemcik@fjfi.cvut.cz}

\abstract{We present for the first time a revised study of 
charmonium production in nuclear ultra-peripheral collisions (UPC) 
based on a rigorous Green function formalism. 
This formalism allows for the proper incorporation of the effects of color transparency, as well as the quantum coherence inherent 
in the higher twist quark shadowing related to the $Q\bar Q$ 
Fock component of the photon. The significance of this effect 
gradually decreases towards forward and/or backward rapidities. 
In the LHC kinematic region we additionally incorporate within 
the same formalism the leading twist gluon shadowing corrections
related to higher multi-gluon photon fluctuations. They represent 
a dominant source of nuclear phenomena in the mid-rapidity region. 
Model predictions for the rapidity distributions $d\sigma/dy$ 
are in good agreement with available UPC data on coherent 
charmonium production at RHIC and the LHC. They can also be 
verified by future measurements at the LHC, as well as at EIC.}

\FullConference{%
  42$^{\rm nd}$ International Conference on High Energy Physics (ICHEP24),\\
  18-24 July 2024\\
  Prague, Czech Republic
}

\def\Jpsi{J\!/\!\psi}
\def\psip{\psi^{\,\prime}}
\def\TeV{\,\mbox{TeV}}
\def\GeV{\,\mbox{GeV}}

\def\sqq{\sigma_{Q\bar Q}}
\def\lsim{\mathrel{\rlap{\lower4pt\hbox{\hskip1pt$\sim$}}
     \raise1pt\hbox{$<$}}}         
 \def\gsim{\mathrel{\rlap{\lower4pt\hbox{\hskip1pt$\sim$}}
     \raise1pt\hbox{$>$}}}         

\begin{document}
\maketitle

\section{Introduction}

The study of heavy quarkonium production 
in heavy-ion collisions can help us to extend our understanding of the QCD dynamics, as well as manifestations of various nuclear effects, such as quantum coherence,
color transparency, gluon shadowing
and gluon saturation. The production of charmonia in photo-nuclear reactions ($Q^2\approx 0$) is recently intensively studied in ultra-peripheral Pb-Pb collisions at the 
Large Hadron Collider (LHC) and in Au-Au UPC at 
Relativistic Heavy Ion collider
(RHIC). Here, the gluon shadowing (GS) corrections at LHC energies and  
the reduced effects of quantum coherence in the forward/backward rapidity region 
are often neglected or incorporated inaccurately in 
most of the current calculations.
In this contribution, we refer on our study of coherent (elastic)
photoproduction of charmonia, $\gamma A\to V A$ [V = $\Jpsi(1S), \psip(2S)$],
in UPC, 
extended to charmonium electroproduction, $\gamma^*A\to V A$, in expected 
Electron-Ion-Collider (EIC) kinematic region
within the light-front (LF) color dipole approach. We focus on the proper treatment of reduced effects of the coherence length using the Green function technique.

\section{
Green function formalism for the coherent photo-nuclear reaction
}

The cross section for photo-nuclear reaction in a heavy-ion UPC,
derived in the one-photon-exchange approximation and
expressed in the rest frame of the target nucleus $A$ \cite{Bertulani:2005ru}, is defined as follows:
%
\begin{equation}
  k\frac{d\sigma}{dk} = \int\,d^2\tau \int\,d^2b \,\,
  n(k,\vec b-\vec\tau,y)\, 
  \frac{d^2
  \sigma_A(s,b)}{d^2b}
  ~~ + ~~
  \Bigl \{ y\rightarrow -y \Bigr \}
  \,.
\label{cs-upc}
\end{equation}
%
%
Here the rapidity variable 
$y = \ln \bigl [s / (M_V \sqrt{s_N}~) \bigr] \approx \ln\bigl [(2 k M_N + M_N^2) / (M_V \sqrt{s_N}~)\bigr ]$ 
with $M_N$ and $M_V$ being the nucleon and vector meson mass, respectively, 
$\sqrt{s_N}$ is the collision energy and $k$ is the photon energy 
related to the square of the photon-nucleon center-of-mass (c.m.) energy 
$W^2 = 2 M_N k + M_N^2 - Q^2\approx s - Q^2$,
where $s$ is the photon-nucleon c.m. energy squared. The variable $\vec\tau$ is the relative impact parameter of a nuclear collision and $\vec{b}$ is the impact 
parameter of the photon-nucleon collision relative to the center of one of the nuclei. It holds that $\tau > 2 R_A$, for a UPC of identical nuclei with the nuclear radius $R_A$. The variable $n(k,\vec b)$ represents the photon flux induced by a projectile nucleus. The corresponding formula for the coherent
production cross section has the following form \cite{Kopeliovich:2001xj,Obertova:2024prd}
%
\begin{eqnarray}
  \label{sigcoh}
   \frac{d^2\sigma_{A}^{coh}(s,b)}{d^2b} 
  = 
  \frac{1}{4}\,
  \Bigl | \int\limits_{-\infty}^{\infty}\,dz\,\rho_{A}({b},z)\,
         H_{1}(s,{b},z)\,\Bigr |^2\, ,
\end{eqnarray}
where $\rho_A(b,z)$ is the nuclear density distribution employed in the realistic Wood-Saxon form. The function $H_{1}(s,b,z)$ 
can be expressed within a rigorous path integral technique as
\begin{eqnarray}\label{f1}
\!\!\!\!\!\!\!\!
H_1(s,b,z) \!\!\! &=&\!\!\!\!\!
\int_0^1 \!\!d\alpha
\int \!\! d^{2} r_{1}\,d^{2} r_{2}\,
\Psi^{*}_{V}(\vec r_{2},\alpha)\,
G_{Q\bar Q}(z^{\prime}\!\!\to\!\infty,\vec r_{2};z,\vec r_{1})\,
\sigma_{Q\bar Q}(r_{1},s)\,
\Psi_{\gamma(\gamma^*)}(\vec r_{1},
\alpha)\,.
 \end{eqnarray}
Here, 
$\Psi_V(\vec r,\alpha)$ is the LF wave function for heavy quarkonium, obtained by the Terent'ev boosting prescription~\cite{Terentev:1976jk} and including the correction due to the Melosh spin rotation effect~\cite{Melosh:1974cu}, and
$\Psi_{\gamma(\gamma^*)}(\vec r,\alpha)$ is the LF distribution 
of the $Q\bar Q$ Fock component of the quasi-real (transversely polarized) or virtual photon.
Variable $\vec{r}$ is the transverse separation of the $Q\bar Q$ 
fluctuation (dipole) and $\alpha = p_Q^+/p_{\gamma}^+$ is 
the boost-invariant fraction of the photon momentum carried by a heavy quark (or antiquark). The dipole cross section $\sigma_{Q \bar{Q}}(r,s)$ in Eq.~(\ref{f1}) describes the interaction of the 
$Q\bar Q$ dipole with the nucleon target. The Green function $G_{Q\bar Q}(z^{\prime}\to\infty,\vec r_{2};z,\vec r_{1})$, which describes the evolution of a $Q\bar{Q}$ pair between points $\vec{r}_1, z$ and $\vec{r_2}, z^{\prime}$, satisfies
the two-dimensional Schr\"odinger equation
\cite{Kopeliovich:1999am}
\begin{equation}
i\frac{d}{dz_2}\,G_{Q\bar Q}(z_2,\vec r_2;z_1,\vec r_1)=
\left[\frac{\eta^{2} - \Delta_{r_{2}}}{2\,k\,\alpha\,(1-\alpha)}
+V_{Q\bar Q}(z_2,\vec r_2,\alpha)\right]
G_{Q\bar Q}(z_2,\vec r_2;z_1,\vec r_1)\,,
\label{schroedinger}
 \end{equation}
where $\eta^2 = m_Q^2 + \alpha (1-\alpha) Q^2$, $m_Q$ is the heavy quark mass and Laplacian $\Delta_{r_2}$ acts on the coordinate $r_2$. Important ingredient of the Schr\"odinger equation is the complex $Q\bar{Q}$ potential in the LF frame, $V_{Q\bar Q}(z_2,\vec r_2,\alpha)$. The imaginary part of the LF potential controls the attenuation of the $Q\bar Q$ pair in the medium and
the corresponding real part describes the interaction between the $Q$ and $\bar{Q}$. The values of the real part of the LF potential for arbitrary model of $Q-\bar{Q}$ interaction are obtained numerically, as described in Ref.~\cite{Obertova:2024prd}, by solving the LF Schr\"{o}dinger equation for $\Psi_{V}(\vec r_{2},\alpha)$.

The kinetic term in Eq.~\eqref{schroedinger} includes the coherence length (CL), $l_c=1/q_L$, with $q_L$ being the longitudinal momentum transfer defined as 
\begin{equation}
q_L(z)
=
\frac{M_{Q\bar Q}^2(z) + Q^2}{2\,k} 
\Rightarrow
\frac{\eta^2 - \Delta_{r_{2}} }{2\,k\,\alpha(1-\alpha)}\,,
\end{equation}
where $M^2_{Q\bar{Q}}=(m_Q^2 + q_T^2) /\alpha (1-\alpha)$ is the $Q\bar{Q}$ effective mass and $q_T^2 \Rightarrow -\Delta_r$ is the quark transverse momentum. The condition $l_c \geq R_A$ controls the onset of the quantum coherence effect 
related to $Q\bar Q$ Fock component of the photon and leading to
higher twist quark shadowing. 
Another effect, known as the color transparency (CT),
is responsible for the final state absorption of produced quarkonium \cite{Kopeliovich:2001xj,Obertova:2024prd}. The onset of CT is controlled by the 
transverse size evolution of a $Q\bar{Q}$ pair propagating through the medium and is also 
incorporated in the Green function formalism.

In the electroproduction of heavy quarkonia 
at EIC energies, we evaluate the total cross section for quarkonium production on nuclear target as follows,
%
\begin{equation}
\sigma_{\gamma^*A\to\Jpsi A}(s,b) = \int d^2 b ~~\frac{d^2\sigma_A^{coh}(s,b)}{d^2 b}\,. 
\label{sigtot}
\end{equation}
Then, we express the nucleus-to-nucleon ratio ({\sl{nuclear transparency}}) as 
%
\begin{eqnarray}\label{eq:RA}
R_A^{coh}(\Jpsi)  = 
\frac{
\sigma_{\gamma^*A\to\Jpsi A}}{A ~\sigma_{\gamma^*N\to\Jpsi N}}
\end{eqnarray}
where $\sigma_{\gamma^*N\to V N}$ is the total cross section for $J/\Psi$ production on a nucleon target.
More details about the formalism can be found in Ref.~\cite{Obertova:2024prd}.

We also include in Eq.~(\ref{f1}) a small correction 
due to the real part of the $\gamma^* N\to \Jpsi N$ amplitude by
performing the following replacement 
\cite{Bronzan:1974jh,Nemchik:1996cw,Forshaw:2003ki},
%
\begin{equation}
\sqq(r,s)
\Rightarrow
\sqq(r,s)
\,
\left(1 - i\,\frac{\pi}{2}\,
\Lambda
\right),
\qquad\qquad
\Lambda = 
\frac
{\partial
 \,\ln\,{\sqq(r,s)}}
{\partial\,\ln s}
\, .
  \label{re/im}
\end{equation}
%

Another phenomenon that affects heavy quarkonium production in photo-nuclear reactions in UPC is gluon shadowing. The effect of GS has to be additionally included as a shadowing correction corresponding to higher Fock components of the photon containing gluons. It was incorporated as
a reduction of $\sqq(r,s)$ in nuclear reactions with respect 
to processes on the nucleon
\cite{Kopeliovich:2001ee},
%
\begin{equation}\label{eq:dipole:gs:replace}
  \sqq(r,x) \Rightarrow \sigma_{Q\bar{Q}}(r,x) \cdot R_G(x,b)\,.
\end{equation}
%
The correction factor $R_G(x,b)$ was calculated within the Green function formalism as a function of the nuclear impact parameter $b$ and the Bjorken variable $x$.

\section{Results}

We calculated the coherent charmonium photoproduction in UPC at LHC energies according 
to Eq.~(\ref{cs-upc}), as well as the electroproduction {at} EIC {energies} 
using Eqs.~(\ref{sigtot}) and (\ref{eq:RA}).
\begin{figure}
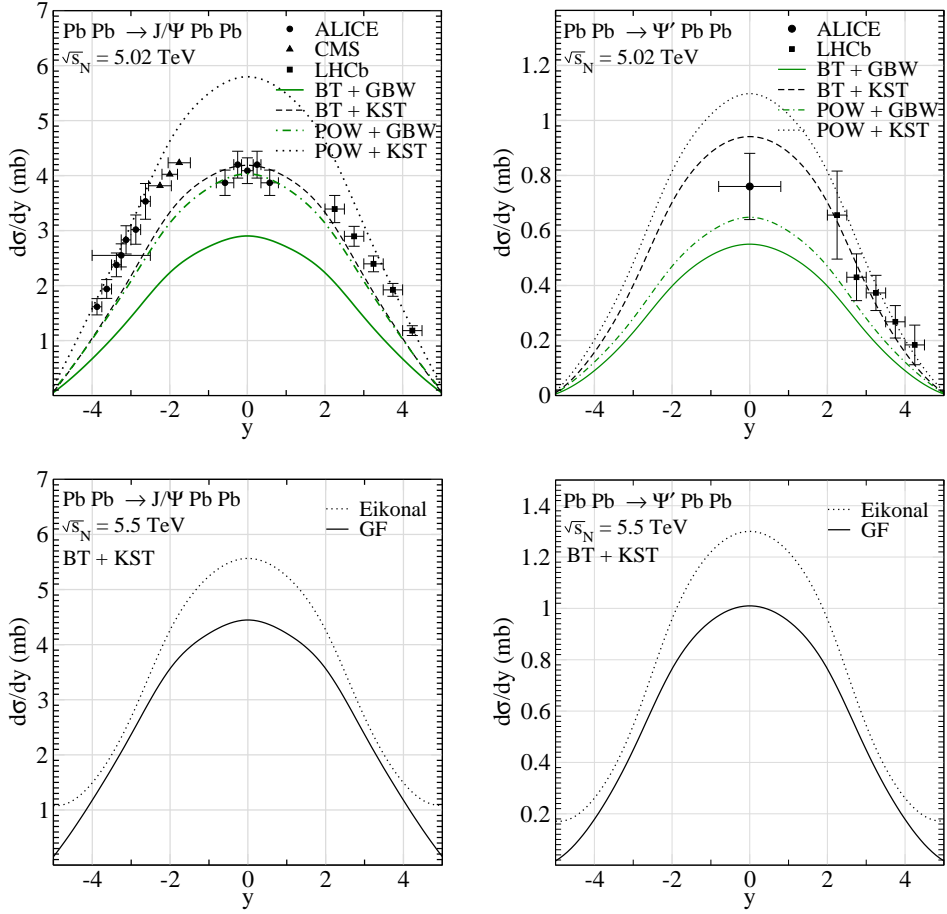

 \centering
 \includegraphics[width=0.38\textwidth]{dsdyGF_psi1S_5020Pb_Re+ImV_newGS.eps}\hspace{15pt} 
 \includegraphics[width=0.395\textwidth]{dsdyGF_psi2S_5020Pb_Re+ImV_newGS.eps}
 \\
 \vspace{10pt}
 \includegraphics[width=0.38\textwidth]{dsdyGF_psi1S_5500Pb_BT_KSTr_eikonal_GF.eps}\hspace{15pt} 
 \includegraphics[width=0.395\textwidth]{dsdyGF_psi2S_5500Pb_BT_KSTr_Re+ImV_newGS.eps}
 \caption{Rapidity distributions of coherent cross section for photoproduction of $J/\psi$ (left) and $\psi^{\prime}$ (right) in UPC at LHC energies $\sqrt{s_N} = 5.02\,\TeV$ (top panel) and 5.5 TeV (bottom panel). The data for $d\sigma/dy$ from  ALICE \cite{Acharya:2019vlb}, LHCb \cite{LHCb:2018ofh,LHCb:2022ahs} and CMS \cite{CMS:2023snh} Collaborations are shown for comparison. Figure is adopted from Ref.~\cite{Obertova:2024prd}.}
 \label{fig:1}
\end{figure}
In Figure~\ref{fig:1}, top panels, we present comparison of our model predictions 
for the rapidity distributions $d\sigma/dy$ of coherent $\Jpsi(1S)$ (left panel) 
and $\psip(2S)$ (right panel) photoproduction in UPC at c.m. collision energy $\sqrt{s_N} = 5.02\,\TeV$ with available data.
Our results were obtained for charmonium wave functions 
generated by two $c$-$\bar c$ potential models, 
POW 
\cite{Martin:1980jx,Barik:1980ai}
(dotted and dot-dashed lines) and BT 
\cite{Buchmuller:1980su}
(solid and dashed lines).
For the dipole cross sections $\sqq$ we employed the GBW~\cite{Golec-Biernat:2017lfv} (solid and dot-dashed lines) 
and KST~\cite{Kopeliovich:2001xj} (dashed and dotted lines) parametrizations. 
Our predictions based on the Green function formalism manifest a rather good agreement with data. Bottom panels of Figure~\ref{fig:1} show the prediction for rapidity distributions of coherent charmonium cross section at larger $\sqrt{s_N} = 5.5\,\TeV$, which is planned to be measured at the LHC. Here, solid lines represent
calculations within a rigorous path integral technique.
The dotted lines are based on the Eikonal approximation ($l_c \gg R_A$ in Eq.~\eqref{f1}) for production cross section 
and without GS corrections. The difference between solid and dotted lines at midrapidities illustrates the net GS effect. At forward/backward rapidities, the GS vanishes and the difference between the two lines
is caused by the reduced effects
of quantum coherence when $l_c\lsim R_A$.

\begin{figure}
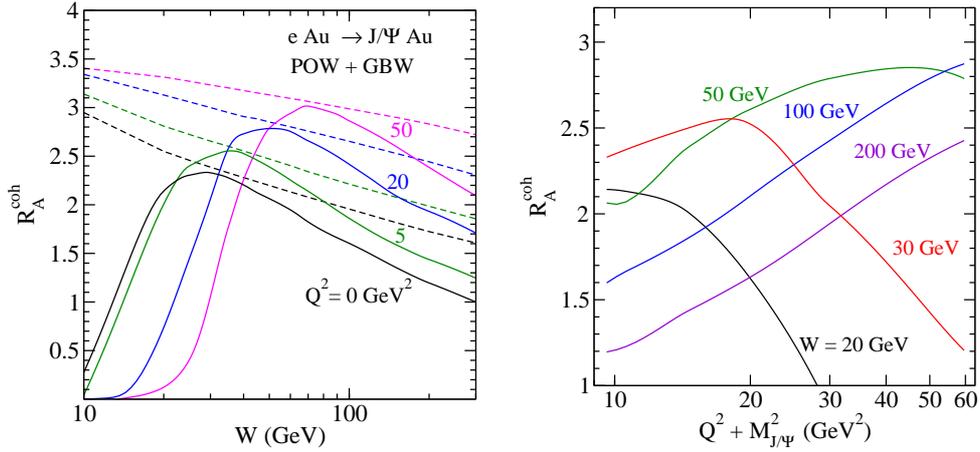

 \centering
 \includegraphics[width=0.41\textwidth]{RcohGF_psi1S_Au_Pow_GBWnew_1.eps} \hspace{10pt}
 \includegraphics[width=0.4\textwidth]{RcohGF_Q2+MV2_psi1S_Au_Pow_GBWnew_new.eps}
 \caption{Ratio $R_A^{coh}$ for $J/\psi$ coherent production on the gold target as a function of c.m. energy $W$ at 
several fixed values of photon virtuality $Q^2 = 0$, 5, 20 and 50 GeV$^2$ (left panel). Right panel shows the ratio $R_A^{coh}$ as a function of $Q^2+M_{J/\Psi}^2$ at several fixed values of $W = 20$, 30, 50, 100 and 200$\,\GeV$. Figure is adopted from Ref.~\cite{Obertova:2024prd}.}
\label{fig:2}
\end{figure}

In Figure~\ref{fig:2}, we present the ratio $R_A^{\rm coh}$ from Eq.~\eqref{eq:RA}
as a function of the c.m. energy $W$ (left panel) and as a function of $Q^2+M_{J/\Psi}^2$ (right panel) for electroproduction of $\Jpsi(1S)$ on a gold target. The values of $R_A^{coh}$ were obtained using the charmonium LF wave function generated by the POW $c$-$\bar c$ interaction potential and the GBW model for $\sqq$ fitted to data for photon virtualities $Q^2\leq50$~GeV$^2$~\cite{Golec-Biernat:2017lfv}.
In the left panel of Figure~\ref{fig:2}, our results show significant leading twist corrections, rising with the photon energy, as differences between solid (the Green function formalism with GS) and dashed (Eikonal approximation without GS) lines at large $W$. At small photon energies, up to the position of maximal $R_A^{\rm coh}$ values, the difference is caused by the effect of reduced coherence length, as $l_c\lsim R_A$. 
Right panel of Figure~\ref{fig:2} represents the $Q^2+M_{J/\Psi}^2$-behavior of $R_A^{\rm coh}$ at different values of photon energy $W$. At large $W= 100$ and $200$ GeV, CL is long enough, $l_c > R_A$, to neglect its variation with $Q^2$
and the rise of $R_A^{\rm coh}$ with $Q^2$ is a net manifestation of CT. The effect of reduced CL is visible at lower energies, $W < 50$ GeV, as decrease of $R_A^{\rm coh}$ with $Q^2$.

Our predictions for the onset of reduced coherence effects and gluon shadowing can be tested 
by future experiments at EIC.
The proper treatment of the shadowing and absorption effects in the nuclear medium within a rigorous path
integral technique may be crucial for the future conclusive evidence of expected gluon saturation effects
at large energies.

\begin{acknowledgments}
The work of J.N. was partially supported by the Slovak Funding Agency, Grant No. 2/0020/22.
Computational resources were provided by the e-INFRA CZ project (ID:90254),
supported by the Ministry of Education, Youth and Sports of the Czech Republic.
\end{acknowledgments}



\end{document}